# Photoluminescence modulation of ZnO via coupling with the surface plasmon resonance of gold nanoparticles

Dongyan Zhang,[1,a)] Hiroki Ushita,[1] Pangpang Wang,[2] Changwook Park,[1,3] Ri-ichi Murakami,[1] Sheng-chun Yang,[4] Xiaoping Song[4]

[1]Graduate School of Advanced Technology and Science, the University of Tokushima, Tokushima 770-8506, Japan

[2]Institute for Materials Chemistry and Engineering, Kyushu University, Fukuoka 812-8581, Japan

[3]Composite Laboratory, Korea Maritime Univerity, Busan, 606-791, South Korea

[4]MOE Key Laboratory for Nonequilibrium synthesis and Modulation of Condensed Matter, Xi'an Jiaotong University, Xi'an 710049, People's Republic of China

**ABSTRACT**

In this letter, we study how coupling between AuNPs and ZnO thin films affects their emission properties. The emission intensity of ZnO thin films changes when $Al_2O_3$ spacer layer of different thickness are included in ZnO/Au films, consistent with theoretical predictions. The emission properties are also controlled using the polarization of the excitation source. Emission properties depended on the polarization of the excitation source

---

a) Author to whom correspondence should be addressed. E-mail: zhang@tokushima-u.ac.jp

because of the surface plasmon resonance of AuNPs. The photoluminescence anisotropy of these systems shows that enhanced photoluminescence can be achieved through coupling of the emission from ZnO with the surface plasmon resonance of AuNPs.

Metal nanoparticles (NPs) have the ability to localize and strongly enhance an incident electromagnetic field because of surface plasmon polaritons.[1, 2] This property has been experimentally demonstrated and used in several applications such as surface enhanced Raman scattering,[3, 4] extraordinary transmission,[5, 6] near-field photopolymerization,[7, 8] and metal-enhanced fluorescence.[9, 10] The enhancement of fluorescence strongly depends on the size and shape of the metal NPs,[11, 12] and especially their surface plasmon properties.[13, 14] Recently, considerable attention has been paid to increasing luminescence using hybrid systems composed of a metal nanostructure and fluorophores. Nearby conducting metallic NPs can modify the interface charge density wave[6] so that it increases or decreases the incident electric field, which affects the emission from the fluorophores. Both enhancement and quenching of photoluminescence (PL) have been observed in experimental results.[14-17] However, the mechanism of increasing and decreasing the intensity of photoluminescence is still debated. The effect of the photoluminescence modulation via coupling with the surface plasmon resonance of NPs depends on the competition between the enhancement of the local excitation field and modification of the rates of the radiative or nonradiative decay.[18, 19] The configuration of hybrid systems containing a metal nanostructure and fluorophores plays an important role in the quenching and enhancement mechanisms.

The theory of fluorescence enhancement and quenching by surface plasmon polaritons has been developed since the 1980s.[20-22] Lakowicz[23, 24] investigated the effects of different silver nanostructures on the emission properties of fluorophores, and also the dependence of the transitions of fluorophores on their distance from a metallic surface. Plain[25] had studied the influence of local surface plasmon resonance (LSPR) of gold (Au) NPs on the emission wavelength of quantum dots. However, the principles underlying these opposing observations remain unclarified. In particular, the influence of surface plasmon polaritons on the emission maximum of luminescent species is unclear.[26, 27]

In this letter, we study the properties of AuNPs coupled with ZnO thin films. The emission intensity increases exponentially as the AuNPs are separated from ZnO. We also investigated the influence of polarized excitation on the emission properties of these systems because PL properties are strongly affected by the polarization of the excitation source. In these systems, PL anisotropies are affected when a local incident electromagnetic field is enhanced. This investigation demonstrates a method of PL modulation *via* surface plasmon resonance.

AuNPs were fabricated via self-assembly at a pentanol-water interface, as previously reported.[28] Fig. 1a and b show TEM images of a monolayer film of AuNPs. AuNPs with a diameter of approximately 50 nm were assembled into nanonetworks. Then, $Al_2O_3$ and ZnO were deposited as a spacer layer and luminescent species, respectively, by dc magnetron sputtering (NACHI, SP-1530-1, Japan) at room temperature. The ZnO layer was approximately 500 nm thick, and the $Al_2O_3$ spacer layer was varied from 0 to 25 nm thick.

Fig. 1c and d show the cross-sectional SEM images of a ZnO/Al$_2$O$_3$ (25 nm)/AuNP film on a glass substrate. The X-ray diffraction analysis for ZnO deposited on Glass, Au NPs/Glass, and Al2O3/Au NPs/Glass were performed, as shown in Figure 1e. The FWHM (full width half maximum) of ZnO (002) peaks indicate that the ZnO deposited on Glass substrate show the best crystal quality, because the Au NPs/Glass and Al$_2$O$_3$/Au NPs/Glass have a worse surface roughness as shown in Fig. 1 a, b and d. This can exclude that the enhancement of photoluminescence are attributed to improvement of ZnO crystal quality.

It is well known that the dispersion relation of surface plasmon lies to the right of the light line.[29] Here, AuNP network monolayer films were fabricated to enhance the local incident electromagnetic field. The interspaces in the nanonetworks increased the wave vector of the incident light to satisfy the dispersion relations of surface plasmon polaritons. Because of surface plasmon resonance, a peak in the visible range should be observed in extinction spectra. The extinction spectra of AuNPs with different surrounding dielectric (air, Al$_2$O$_3$, and ZnO) are shown in Fig. 2. Changing the surrounding dielectric from air to Al$_2$O$_3$ to ZnO caused the LSPR wavelength to increase 530 nm to 545 nm to 552 nm. According to Mie theory,[30] the extinction cross section of NPs can be described by the following relationship:[31]

$$\sigma_{ext} = \frac{9V\epsilon_m^{3/2}}{c} \cdot \frac{\omega\varepsilon_2(\omega)}{[\varepsilon_1(\omega)+2\epsilon_m]^2+\varepsilon_2^2(\omega)} \qquad (1)$$

where $V$ is the spherical particle volume, $c$ the velocity of light, $\omega$ the angular frequency of incident light, and $\varepsilon_m$ the dielectric constant of the surrounding medium. $\varepsilon_1$ and $\varepsilon_2$ are the real and imaginary components of the dielectric function for the AuNPs. The wavelength of the maximum absorption could be estimated when the $\varepsilon_1(\omega)+2\varepsilon_m=0$. Assuming the Drude model is applicable in this case, Eq (1) could be reduced as follows:

$$2\epsilon_m = \frac{\omega_p^4}{\omega^2-\gamma^2} - 1 \qquad (2)$$

where $\omega_p$ is the bulk plasmon frequency and $\gamma$ the size limit relaxation time for Au. Eq (2) gives the relationship between the LSPR wavelength and surrounding dielectrics. An increase in the dielectric constant of the surrounding causes the LSPR to exhibit a red-shift, which was verified in our experiment (Fig. 2). The extinction maximum for a ZnO or $Al_2O_3$ surrounding is at longer wavelength than that for air. Furthermore, the extinction maximum is red-shifted with respect to the PL wavelength of ZnO. Even though ZnO is coated on $Al_2O_3$/Au, the extinction maximum is located to the long-wavelength side of that of AuNPs (Curve 4, Fig. 2). This is a requirement for modulation of PL because the relative position between emission and extinction maxima determines whether quenching or enhancement of PL is observed.[25] It also implies that the change of PL intensity is not attributed to the shift of LSPR. Modification of PL properties was instead achieved by embedding an $Al_2O_3$ spacer layer or adjusting the polarization properties of the excitation source.

The 500 nm-thick ZnO film (Fig. 1c) was used as a luminescent species, and was excited by incident light with a wavelength of 325 nm. Fig. 3 shows the PL properties of ZnO/$Al_2O_3$/Au with $Al_2O_3$ layers of different thickness. The PL spectrum of ZnO is included as a reference.

Importantly, the intensity of band-edge emission was greatly affected by the thickness of the spacer layer. The observed enhanced emission is attributed to surface plasmon. When ZnO was brought into contact with the AuNPs, the surface plasmons of AuNPs increased the local incident field on the fluorophores, causing the emission intensity to increase. As shown in Fig. 3, the emission of ZnO/Al$_2$O$_3$/Au was stronger than that of bare ZnO. However, other mechanisms also need to be considered, such as quenching and space charge. The quenching is caused by energy transfer to the metal, which would decrease the emission of ZnO. This quenching was attributed to damping of the dipole oscillations by the nearby AuNPs. The space charge would affect the magnitude of surface plasmon, which would also affect the nearby ZnO emission[32]. The surface plasmon, quenching and space charge jointly decided the decrease and increase of the photoluminescence of AuNPs contacted ZnO. Such competition between quenching and enhancement can be understood using a Jablonski diagram[33]. Absorption of a photon transfers an electron to excited state, and then the electron is relaxed to a sub-stable state through internal conversion. The rate of photon emission from excited ZnO molecules indicates the time it takes for the excited electron to return to the ground state. Other deactivation processes to depopulate the first singlet level by transferring an electron to the ground state are also possible, such as nonradiative decay and quenching. Moreover, the effects of surface plasmons should be taken in account near the AuNPs, which include an increase in the local incident field upon excitation and enhancement of intrinsic radiative.

1　Surface plasmons enhance PL, as shown in Fig. 3a. When an $Al_2O_3$ spacer layer was used to

2　separate the AuNPs and ZnO layer, the emission intensity increased as the thickness of $Al_2O_3$

3　spacer layer increased from none to 25 nm, and then decreased as the thickness further

4　increased (Fig. 3a and b). Taking into account the properties of surface plasmons, its

5　attenuation in a dielectric layer should be much weaker than in a metallic one. This implies

6　that the ZnO fluorophores were still influenced by the enhanced local incident field caused by

7　surface plasmons even when an $Al_2O_3$ spacer layer is present. However, the transfer of

8　energy related to quenching could not penetrate the spacer layer. As a result, as the thickness

9　of the spacer layer increased, the effect of quenching was removed. Notably, the dependence

10　of PL intensity on spacer thickness could be approximately fitted with an exponential

11　function, because the amplitude of surface plasmon resonance decayed exponentially with

12　distance from the surface in the dielectric layer[34], and quenching *via* energy transfer is also

13　depends exponentially on distance.[35] Therefore, the experimental results in Fig. 3 are

14　consistent with surface plasmon theory. The photoluminescence intensity as a function of

15　spacer layer thickness could be jointly decided by the surface plasmon, quenching and space

16　charge, and could be described as the following relationship:

17　$I(x) = I_0 - I_q e^{-\frac{4\pi}{h}\int_0^x 2m[U_0-E]dt} + I_{sp} e^{-\left(\frac{2t}{x}\right)}\rho(E_g) + I_{sc} n_{10} e^{[-(\frac{e\varphi_{n,D}}{kT}+\frac{eF_c x}{kT}+\frac{x^2}{2L_D^2})]}$　　　(3)

18　where $I_0$ is the emission intensity of bare ZnO film, $I_q$, the reduced emission due to quenching

19　by energy transfer to the metal, $I_{sp}$, the enhanced emission due to surface plasmon, $I_{sc}$, the modified emission due to the space charge. The three exponential factors indicate the distance effect on quenching[36], surface plasmon[37], and space charge[38, 39]. The simulation results are represented in Figure 3.

The ability of surface plasmon resonance of AuNPs to enhance the local incident field was investigated using a polarizer to filter the excitation light. It is expected that the excitation properties of the system would vary under these conditions because the angle between the c-axis of ZnO and electric component of incident light is changed. The intensity of emission from ZnO with polarized excitation is shown in Fig. 4b (inset) as a reference. As the polarization angle of excitation of ZnO is changed from 0º to 90º, the emission intensity gradually decreases. The participation of AuNPs should affect the evolution of emission as the polarization angle of excitation light changes. As discussed above, the local incident electromagnetic field was enhanced by surface plasmon. Surface plasmons are an electromagnetic excitation that can be modulated by polarization of incident light;[34] that is, the local incident electromagnetic field can be changed by controlling the polarization of the excitation source. The excitation polarization dependence of the emission intensity from $ZnO/Al_2O_3/Au$ should perform differently to that from ZnO. Fig. 4 shows the PL spectra for $ZnO/Al_2O_3/Au$ with polarized excitation. The polarization angle was defined as the angle between the polarized light source and the vertical direction. Fig. 4b shows the band-edge emission intensity of $ZnO/Al_2O_3/Au$ as a function of the angle of polarized excitation. The band-edge emission intensity increases with polarization angle from 0 to 42°, and then decreases. In comparison, the band-edge emission intensity of ZnO decreases as the polarization angle increases. The variation in emission intensity was related to the coupling between incident light and surface plasmon. In the configuration used for these PL measurements, the electric field component of excitation was parallel to the sample when the polarization angle was 0°. This implies that it was more difficult for surface plasmon to couple with incident light with a 0° polarization than that with a higher angle. As a result, emission intensity at 0° was weaker than that at other polarization angles. The dependence of emission on polarization indicates that it is possible to control the PL properties of NP/fluorophores systems *via* the surface plasmon resonance of AuNPs.

In conclusion, we investigated the PL properties of $ZnO/Al_2O_3/Au$ films. Damping of the dipole oscillations by the AuNPs quenched PL, whereas increasing the local incident field enhanced it. Competition between quenching and enhancement of PL lead to the observed emission intensity in PL spectra. Surface plasmon is the main factor that enhances PL. PL spectra measured under different polarization angles showed that the ability of AuNPs to modify emission intensity depends strongly on the polarization of incident light. Therefore, PL can be controlled by separating the AuNPs and ZnO or by changing the polarization of the excitation source.

1    **ACKNOWLEDGMENT**

2    This work was supported by the National Natural Science Foundation of China (NSFC, Grant

3    No. 51271135). This work was also supported by the Double Degree Program (DDP) of the

4    University of Tokushima.

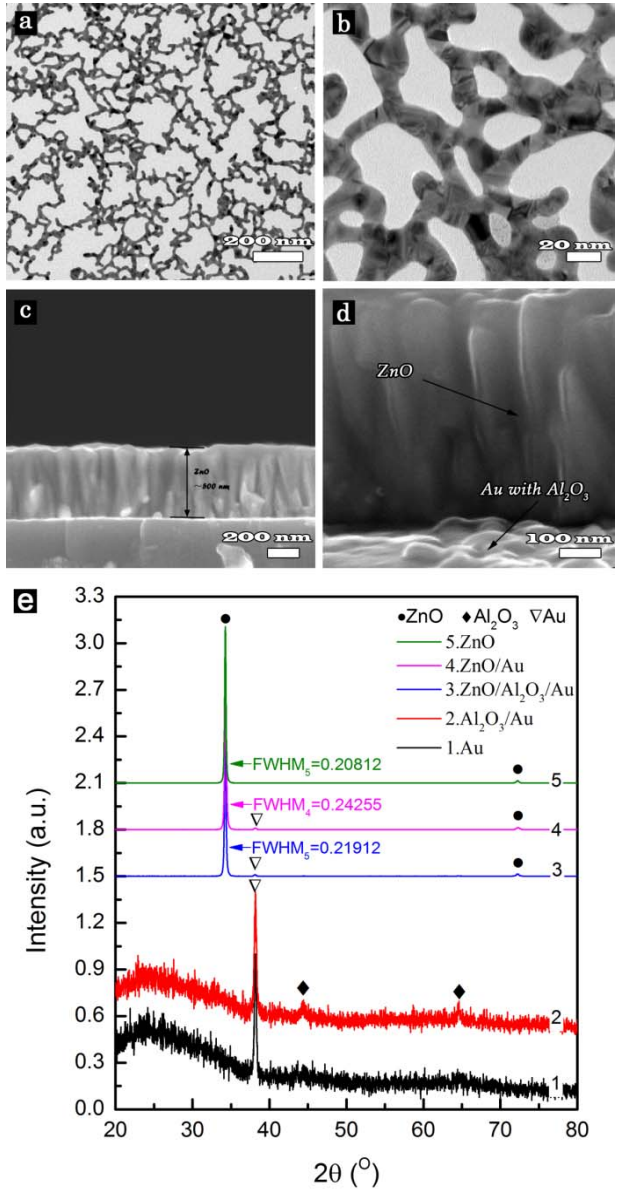

**Figure 1.** TEM images of GNPs and cross-section SEM images of ZnO/Al$_2$O$_3$/Au/Glass multilayer films. (a) TEM image of GNPs networks, (b) Enlarged TEM images, (c) cross-section SEM image of ZnO/Al$_2$O$_3$/Au/Glass, showing approximately 500 nm thickness, (d) Enlarged SEM images, showing the GNPs with coating Al$_2$O$_3$ at the bottom of the multilayer films, (e) XRD patterns for ZnO, ZnO/Au, ZnO/Al$_2$O$_3$/Au, Al$_2$O$_3$/Au, Au.

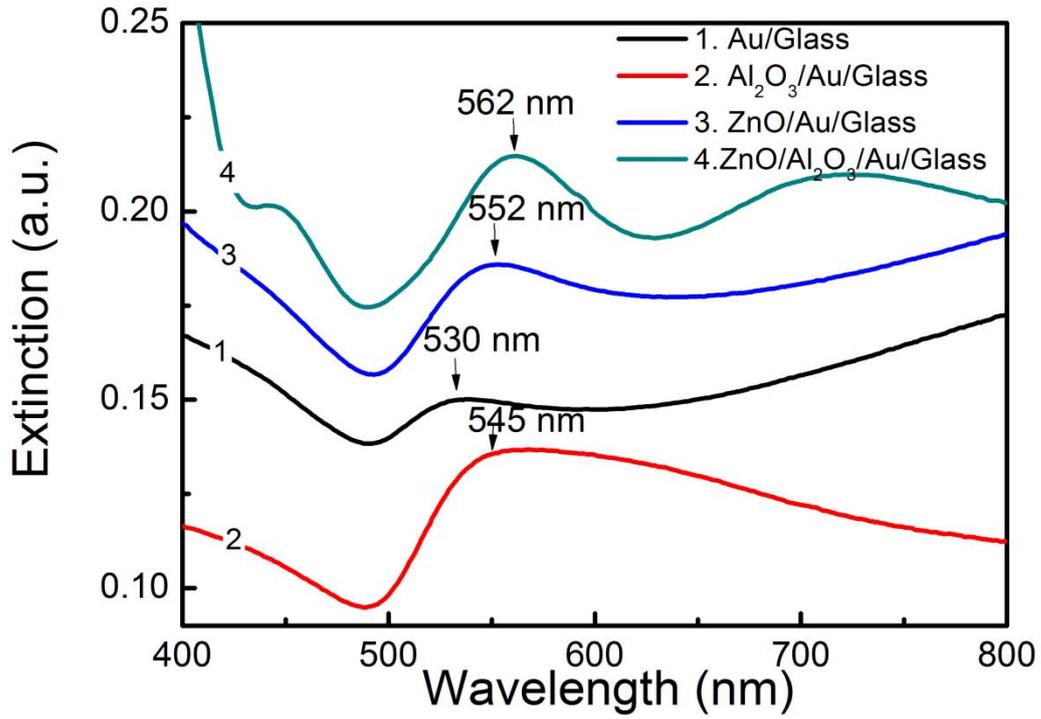

Figure 2. Extinction spectra for Au, Al$_2$O$_3$/Au, and ZnO/Au. The arrows indicate the wavelengths of extinction peaks.

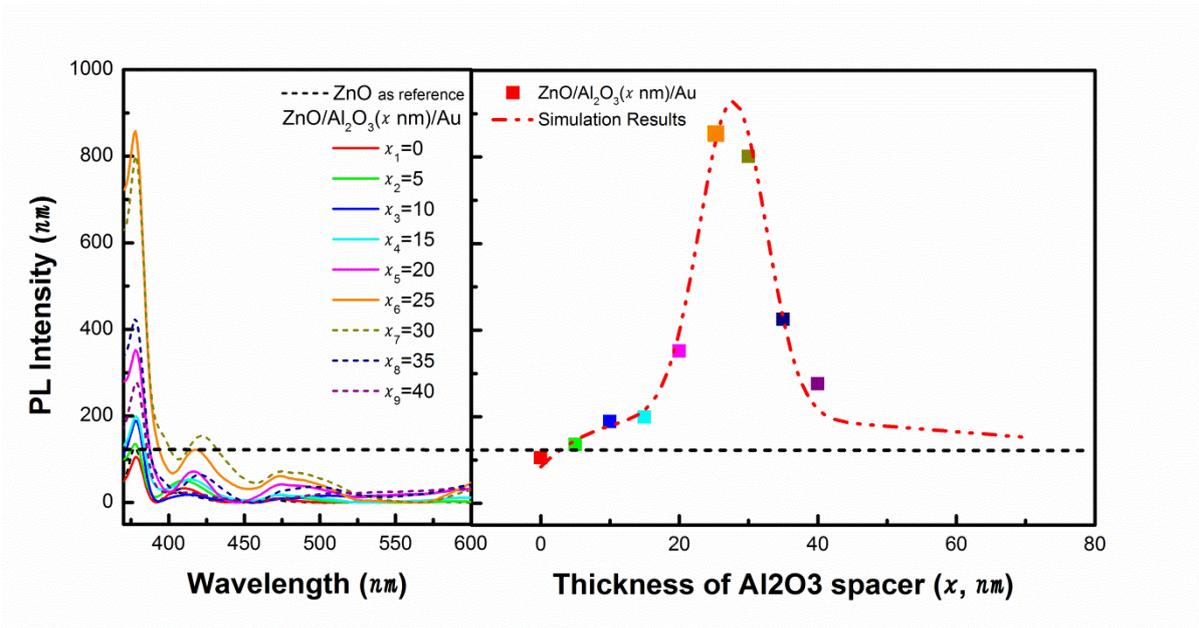

**Figure 3**. (a) PL spectra for $ZnO/Al_2O_3/Au$ with different $Al_2O_3$ thickness. (b) The PL intensity as a function of the $Al_2O_3$ thickness.

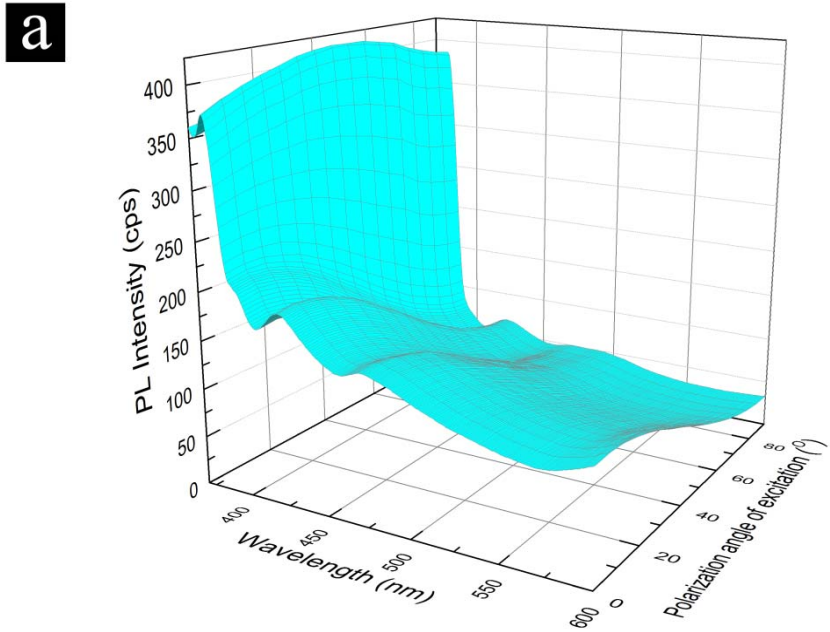

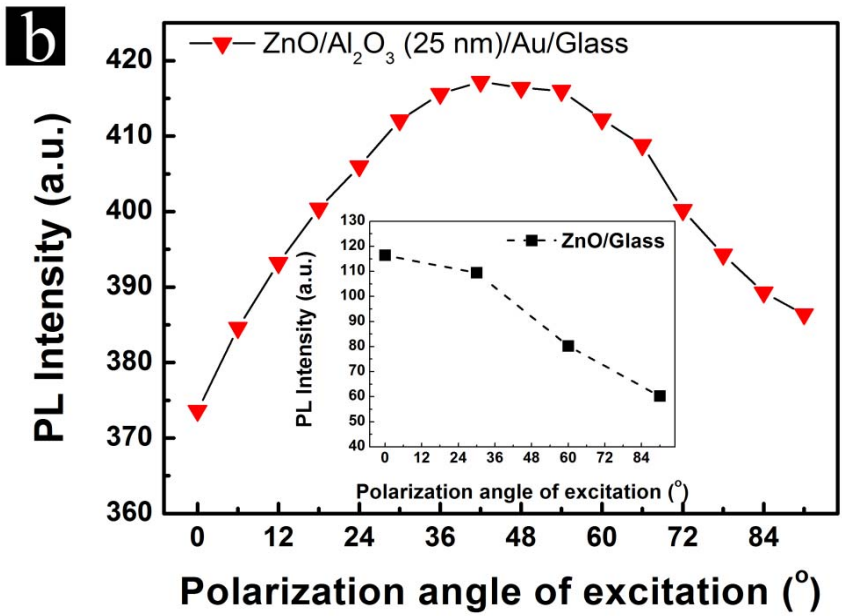

**Figure 4**. PL spectra for ZnO/Al2O3/Au/Glass with a polarization excitation. (a) PL spectra with a different polarization angle of excitation, (b) PL intensity as a function of the polarization angle, inset showing the PL intensity of ZnO dependency on polarization angle as reference.

1    39. A. J. Bennett, Phys Rev B **1** (1), 203-207 (1970).